\documentclass{INTERSPEECH2023}
\usepackage{xcolor}

\interspeechcameraready

\title{Audio-Visual 
Mandarin Electrolaryngeal Speech Voice Conversion}
\name{Yung-Lun Chien$^{1,2}$, Hsin-Hao Chen$^{1,2}$, Ming-Chi Yen$^{2}$, Shu-Wei Tsai$^3$, \\Hsin-Min Wang$^2$, Yu Tsao$^2$, and Tai-Shih Chi$^1$}
\address{
 $^1$ National Yang Ming Chiao Tung University,
$^2$ Academia Sinica\\
$^3$ National Cheng Kung University Hospital}
\email{ajul1230@gmail.com,123ggg3304@gmail.com,ymchiqq@iis.sinica.edu.tw,tsaisuwei@gmail.com,\\whm@iis.sinica.edu.tw,yu.tsao@citi.sinica.edu.tw,tschi@mail.nctu.edu.tw}

\begin{document}

\maketitle
 
\begin{abstract}
Electrolarynx is a commonly used assistive device to help patients with removed vocal cords regain their ability to speak. Although the electrolarynx can generate excitation signals like the vocal cords, the naturalness and intelligibility of electrolaryngeal (EL) speech are very different from those of natural (NL) speech. Many deep-learning-based models have been applied to electrolaryngeal speech voice conversion (ELVC) for converting EL speech to NL speech. In this study, we propose a multimodal voice conversion (VC) model that integrates acoustic and visual information into a unified network. We compared different pre-trained models as visual feature extractors and evaluated the effectiveness of these features in the ELVC task. The experimental results demonstrate that the proposed multimodal VC model outperforms single-modal models in both objective and subjective metrics, suggesting that the integration of visual information can significantly improve the quality of ELVC. 
\end{abstract}

\section{Introduction}
\label{sec:intro}
The ability to speak and communicate is fundamental for human life. However, individuals who undergo laryngectomy lose the ability to produce excitation signals because of the removal of their vocal cords. This loss significantly affects their ability to speak normally, decreasing their overall quality of life. To address this issue, the use of the electrolarynx is the primary method for speech recovery. However, this device often produces a relatively flat fundamental frequency (F0) and generates noise that affects the voice quality, highlighting the need for improved electrolaryngeal (EL) speech techniques.

Voice conversion (VC) is a technique that converts a human voice from a source speaker to target speaker without changing the underlying content. One of the applications of VC is to improve the naturalness and intelligibility of EL speech \cite{doi2011evaluation,yen2021mandarin}; this VC task is called electrolaryngeal speech voice conversion (ELVC). A typical ELVC approach first extracts the acoustic features of EL speech and target natural (NL) speech and then trains a conversion model. When in use, the converted features are synthesized back into a waveform using a vocoder. For frame-based VC, aligning the acoustic features of paired EL and NL speech is critical before training the conversion model. Dynamic time warping (DTW) is the most commonly used algorithm for determining the best alignment path over two feature sequences based on a predefined distance (e.g., the Euclidean distance). However, in ELVC, the DTW algorithm often fails to find the correct alignment path and causes the model to fail in learning the correct conversion function, which seriously affects the performance of ELVC. To address this issue, Liou \textit{et al.} used lip images instead of acoustic features for alignment \cite{liou2021time}. Although this method achieved better ELVC results, it was not the best alignment method. In this study, we explored different alignment methods to improve the performance of ELVC.

In addition to its role in alignment, the lip shape may play an important role in speech signal processing \cite{mcgurk1976hearing}. Although users of the electrolarynx cannot speak normally, their lip movements are similar to those of healthy people. Therefore, the use of lip-shape information to improve the ELVC model is worth studying. Multimodal training methods have been employed in many speech-processing studies \cite{hou2018audio,gao2021visualvoice}, including the VC task \cite{zhou2019multimodal}. In this study, we evaluated different visual feature extractors and determined the best one for the ELVC task. The main contributions of this study are twofold: i) the proposal of a new feature-alignment method suitable for frame-based ELVC, and ii) a novel multimodal VC architecture that uses both acoustic and visual features.

The remainder of this paper is organized as follows. Section 2 introduces the alignment methods, including the traditional and proposed methods. Section 3 introduces different lip-image feature extractors and their uses. Section 4 presents the experimental setup and various objective and subjective evaluations. Finally, Section 5 presents the conclusions of this study and directions for future research.

\section{Alignment methods}
\label{sec:alignment}
In this section, we will introduce previous and our alignment methods for ELVC.

\subsection{Previous alignment methods}

\begin{figure}[t]
    \centering
    \scalebox{1.0}{
    \includegraphics[width=1.0\columnwidth]{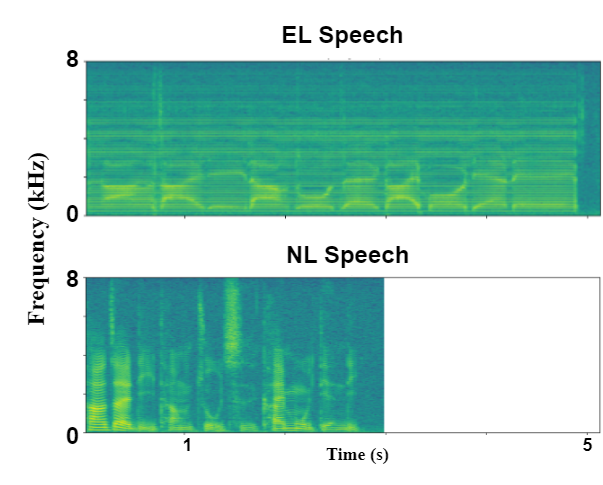}}
    \caption{Spectrogram plots of EL speech and NL speech.}
    \label{fig:ELNL}
\end{figure}

As shown in Fig.~\ref{fig:ELNL}, EL speech is generally longer than NL speech, even with the same linguistic content. Differences in the speech length can cause distortion of NL speech owing to the stretching of length during alignment. In addition to the very different acoustic properties of EL and NL speech, the length difference is one of the key challenges in aligning these two types of speech.

As a baseline, we used the WORLD vocoder \cite{morise2016world} to decompose EL and NL speech into acoustic features, such as mel-cepstral coefficients (MCC). Subsequently, an alignment was performed based on the DTW algorithm using MCC. This method is referred to as DTW-MCC. The path calculation is based on the mel-cepstral distortion (MCD).

Liou \textit{et al.} used the lip images of EL speech and NL speech to align both \cite{liou2021time}. This approach involves first obtaining 20 lip landmarks using the dlib library \cite{king2009dlib}, relocating the coordinates according to their centroid, and then calculating the Euclidean distance between the source and target landmark sets. Although the DTW-lip-landmark method was shown to outperform the DTW-MCC method in the ELVC task in \cite{liou2021time}, the room for improvement exists.

\subsection{Proposed alignment method}

To address the misalignment caused by the difference in length of EL and NL speech, we applied the waveform similarity overlap-and-add (WSOLA) algorithm \cite{verhelst1993overlap}, which is a time-scale modification method that can adjust the speed of speech while preserving F0. Specifically, we used WSOLA to adjust the length of NL speech to match that of the EL speech, thereby reducing the distortion caused by the length difference. The modified DTW-MCC method that uses length-adjusted NL speech is referred to as the DTW-WSOLA method. We conducted preliminary listening tests and confirmed that the intelligibility of the NL speech was not compromised after length adjustment.

\section{Multimodal system architecture}

\label{sec:mmsystem}
\begin{figure}[t]
    \centering
    \scalebox{1.0}{
	\includegraphics[width=1.0\columnwidth]{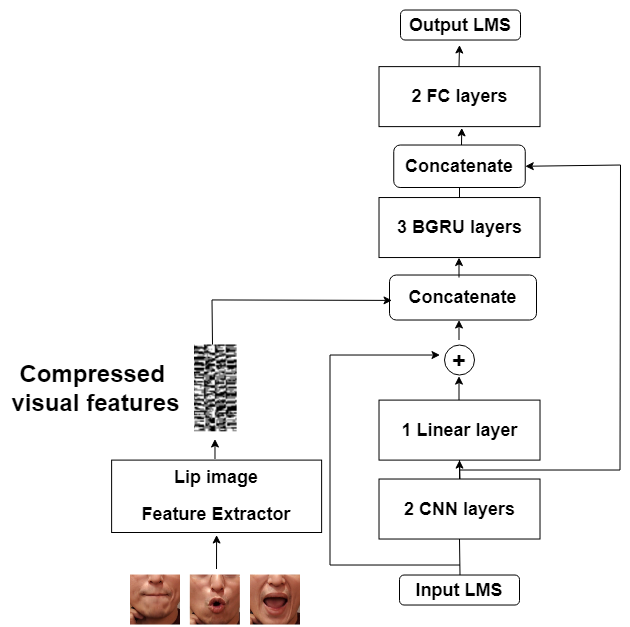}}
	\caption{Overall architecture of the proposed multimodal ELVC system.}
	\label{fig:multi}	
\end{figure}

The overall architecture of the proposed multimodal ELVC system, which consists of a VC model and lip image feature extractor, is illustrated in Fig.~\ref{fig:multi}. The VC model and lip-image feature extractor are described in detail in the following sections.

\subsection{Voice conversion model}
The VC model is implemented based on the CLDNN model proposed in \cite{sainath2015convolutional}. CLDNN has been used in ELVC with satisfactory results in \cite{kobayashi2018electrolaryngeal}. Using the MCC features as the model input, three independent CLDNN models were trained to predict the target speaker’s MCC, aperiodicity (AP), and F0 and unvoiced/voiced (U/V) symbols. To reduce the experimental variability, we changed the input to a logarithmic Mel spectrogram (LMS) and trained a single CLDNN to convert the input LMS into the target LMS. To synthesize the waveform from the LMS, we used parallel WaveGAN \cite{yamamoto2020parallel} as the vocoder in our experiments.

\subsection{Lip image feature extractor}

The compressed visual features were obtained using a lip-image feature extractor. The feature extractor can be completely removed during the training phase. The lip-image feature extractors used in this study are described below.

\subsubsection{CNN encoder}
The overall architecture of the CNN-based lip-image feature extractor includes an encoder and a decoder \cite{chuang2020lite}. The encoder consists of three 2D convolutional layers and one linear layer. The decoder architecture is similar to that of the encoder; however, the convolutional layers are replaced by 2D transposed convolutional layers. The CNN-based model was trained in a self-supervised manner by reconstructing input lip images. Then, the lip images were processed by the pre-trained encoder to obtain latent representations of dimension 768, and these representations were used as the visual features for the multimodal VC model.

\subsubsection{Vision Transformer}
Vision Transformer (ViT) \cite{dosovitskiy2020image} is an image classification model with Transformer \cite{vaswani2017attention} as the backbone. We used a pre-trained ViT model\footnote{https://github.com/google-research/vision{\_}transformer} as a lip image feature extractor. The lip images were processed using the ViT model, and 768-dimensional representations of the last hidden layer were used as the visual features for the multimodal VC model.

\subsubsection{AV-HuBERT}
In recent years, many model architectures for self-supervised learning (SSL) have been developed, including AV-HuBERT \cite{shi2022learning}, which inputs both acoustic features and lip images during training. AV-HuBERT enables the model to learn better features through the complementarity of information provided between the two modalities, leading to better results for downstream tasks that utilize lip information. We used a pre-trained AV-HuBERT model \footnote{https://github.com/facebookresearch/av{\_}hubert} as the lip image feature extractor.

When using AV-HuBERT as a feature extractor, it is possible to analyze whether the output of each layer of the transformer encoder is helpful for the ELVC task. Inspired by \cite{hung2022boosting}, a weighted-sum (WS) method was used for the output of each layer to combine the best-fit features. During VC model training, the AV-HuBERT model was fixed, but the weights were learned and updated. To balance the values of the output features of each layer, the output features were normalized and multiplied by the weight values. In our experiments, we compared the performance of the output features of the last hidden layer (LL) with that of the features using the WS method.

\section{Experiments}
\label{sec:experiment}
This section presents the experimental setup, including the data and evaluation metrics, and the experimental results.

\subsection{Datasets and evaluation metrics}

We conducted experiments on the Mandarin parallel ELVC corpus, which was recorded by a doctor imitating a total laryngectomy patient using an electrolaryngeal device. The doctor read each sentence in the phonetically balanced TMHINT \cite{huang2005development} dataset with and without the use of electrolarynx, while the audio and video were simultaneously recorded. We used 288 and 18 utterances as training and test data, respectively. All the speech utterances were sampled at a frequency of 16 kHz. Each speech waveform was converted into an 80-dimensional LMS with a window size of 512 points and frame shift of 160 points. The layer parameters of the CLDNN model architecture in Fig.~\ref{fig:multi} are similar to those in \cite{kobayashi2018electrolaryngeal}, except for the last fully connected layer. Since the input acoustic feature is an 80-dimensional LMS, the number of hidden units in the last fully connected layer is set to 80 to ensure that the input and output dimensions are consistent. The parallel WaveGAN used to synthesize the LMS back into a waveform was trained using the TMSV dataset \cite{chuang2020lite}.

The frame rate of the video was 50 FPS, and we downsampled the frame rate to 25 FPS, such that one image corresponded to four acoustic frames. Lip images were acquired by the lip-image extractor in \cite{martinez2020lipreading} and converted into lip-image features using a lip-image feature extractor. In the experiments, the lip-image feature sequence was aligned with the acoustic frame sequence for model training. The batch size was 16, the learning rate was set to 0.0005, and the Adam optimizer was used.

Three objective metrics were used to evaluate the ELVC systems, including MCD, the syllable error rate (SER) measured by an ASR system trained on the MATBN dataset \cite{wang2005matbn}, and the estimated mean opinion score (MOS) of the pre-trained MOSA-Net \cite{zezario2022deep}\footnote{https://github.com/dhimasryan/MOSA-Net-Cross-Domain}. The SER and predicted MOS values were 7.3\% and 3.052 for NL speech and 82.3\% and 1.556 for EL speech, respectively. These values were considered the upper and lower bounds of the performance of the ELVC models.

\begin{table}[t]
    \caption{Results using different DTW methods.}
    \centering
    \label{tab:DTW}
    \begin{tabular}{cc}
    \hline
    Method & MCD (dB) \\
    \hline
    \textbf{DTW-MCC} & 7.46 \\
    \textbf{DTW-lip-landmark} & 7.21 \\
    \textbf{DTW-WSOLA} & \textbf{6.83} \\
    \hline
    \end{tabular}
\end{table}

\begin{table}[t]
\caption{Results using different lip image feature extractors.}
\centering
\label{tab:FE-obj}
\begin{tabular}{  c|c c c } 
  \hline
  Feature Extractor & MCD (dB) & SER (\%) & MOS \\ 
 \hline
 \textbf{None} & 6.83 & 73 & 1.965  \\
 \textbf{CNN encoder} & 6.81 & 72.6 & 1.972  \\
 \textbf{ViT} & 6.76 & 71.7 & 1.977  \\
 \textbf{AV-HuBERT(LL)} & 6.45 & 69.2 & 2.073  \\
 \textbf{AV-HuBERT(WS)} & \textbf{6.32} & \textbf{66.7} &  \textbf{2.077}  \\
 \hline
 \end{tabular}
\end{table}

\subsection{Experimental results}

Experiments were conducted in two stages. First, the ELVC results obtained using different alignment methods were compared, and the best alignment method for use in subsequent experiments was determined. Subsequently, we compared the ELVC results obtained using different lip-image feature extractors.

\subsubsection{Comparison of alignment methods}

Table \ref{tab:DTW} lists the results obtained by applying different alignment methods to ELVC. The best-performing method was DTW-WSOLA, which stretched the target speech length so that more corresponding acoustic frames were aligned with the EL speech. While this could lead to distortion, it performed better than the DTW-lip-landmark method, which uses lip images for alignment. DTW-WSOLA cannot fully solve the alignment problem caused by the large difference in the acoustic characteristics of EL and NL speech; however, it is much better than other alignment methods. Therefore, DTW-WSOLA was used as the alignment method in subsequent experiments.

\subsubsection{Comparison of visual feature extractors}

Table \ref{tab:FE-obj} lists the results of applying different lip image feature extractors to ELVC. The visual features extracted by the CNN encoder and ViT showed no notable improvement in all three metrics. However, the visual features extracted by AV-HuBERT, both LL and WS, had a significant improvement in MCD, and the WS visual features were more helpful than the LL visual features. Compared with the CNN encoder and ViT, AV-HuBERT used both acoustic features and lip images as model input, which can extract meaningful features and provide more information to better train the conversion model.

\subsubsection{Fine-tuning visual features}

In our previous experiments, we concatenated the visual features extracted using a lip-image feature extractor with the acoustic features and trained a conversion model. In this experiment, we aimed to improve the conversion ability by fine-tuning (FT) the extracted visual features. We fed the extracted visual features to a unidirectional GRU layer and maintained the dimensionality of the features, enabling the model to learn dynamic information between images. The GRU module was trained together with the VC model. Comparing the results in Tables \ref{tab:FE-obj} and \ref{tab:FT-obj}, it is found that the simple FT method can effectively improve the usability of the visual features extracted by all the lip image feature extractors. 

\begin{table}[t]
\caption{Results using different fine-tuning visual features.}
\centering
\label{tab:FT-obj}
\begin{tabular}{c|ccc} 
  \hline
  Method & MCD (dB) & SER (\%) & MOS \\ 
 \hline
 \textbf{CNN encoder+FT} & 6.62 & 70.3 & 2.055  \\
 \textbf{ViT+FT} & 6.56 & 68 & 2.047  \\
 \textbf{AV-HuBERT(WS)+FT} & \textbf{6.28} & \textbf{62.7} &  \textbf{2.113}  \\
 \hline
 \end{tabular}
\end{table}

\begin{table}[t]
    \caption{Subjective evaluation of intelligibility.}
    \centering
    \label{tab:Sub}
    \begin{tabular}{cc}
    \hline
    System & Intelligibility \\ 
    \hline
     \textbf{Audio-only CLDNN} & 2.586  \\
    \textbf{AV-HuBERT(WS)} & 2.951  \\
     \textbf{AV-HuBERT(WS)+FT} & \textbf{3.218}  \\
     \hline
    \end{tabular}
\end{table}

\subsubsection{Subjective evaluation}
For subjective evaluation, an intelligibility test was conducted. During testing, one converted EL speech item was played for each question, and the subjects were asked to rate intelligibility on a scale of 1–5, regardless of the speech quality. The evaluation criteria are as follows: 5 means that every word in the sentence can be understood; 4 means that a few words in the sentence cannot be understood, but it does not affect the understanding of the sentence; 3 means that nearly half of the words in the sentence can be understood, and the content of the sentence can be roughly judged; 2 means that only a few words in the sentence can be understood, but not the whole sentence; and 1 means that the sentence cannot be understood at all.

 Table \ref{tab:Sub} presents the subjective evaluation results of three ELVC systems. The listening test was conducted on 12 untrained but experienced normal hearing subjects. Among them, 8 were male, and 4 were female. The average age of these 12 subjects was 24 years old. For each test sample, participants were not informed which ELVC system was used to generate it.
 We selected 18 speech utterances converted from each ELVC system to conduct the subjective test. Both audio-visual systems (AV-HuBERT(WS) and AV-HuBERT(WS)+FT) using the AV-HuBERT features achieved higher intelligibility than the Audio-only CLDNN system; and the system with fine-tuned visual features (AV-HuBERT(WS)+FT) achieved the best intelligibility. The subjective evaluation results confirm that multimodal learning can help with the ELVC task.

\section{Conclusions and future work}

\label{sec:conclusion}
In this study, we proposed a multimodal ELVC approach. The experimental results show that the quality and intelligibility of converted EL speech can be improved. The features of the SSL models that have been frequently used in recent years also play a pivotal role in our model. In future research, we will attempt to fine-tune the pre-trained AV-HuBERT model to generate more useful features for ELVC. We will also leverage the features of AV-HuBERT to help align EL and NL speech for better ground truth when training the conversion model.

\bibliographystyle{IEEEtran}

\end{document}